\documentclass[12pt]{article}
\usepackage {graphicx}
\newcommand{\dfrac}{\displaystyle\frac}
\newcommand{\nn}{\nonumber}
\newcommand{\lb}{\linebreak}
\newcommand{\q}{\quad}

\title{A STANDARD MODEL\\IN FOLDY-WOUTHUYSEN REPRESENTATION}
\author{V. P. Neznamov}
\begin{document}
 \maketitle
\begin{abstract}

The paper formulates a standard model in the Foldy-Wouthuysen
representation using previously developed approaches as applied to
quantum electrodynamics. The formulation of the theory in the $FW$
representation does not require obligatory interaction of Higgs
bosons with fermions for $SU(2)$ invariance.

In this approach the Higgs boson spectrum is narrowed significantly: the
Higgs bosons are responsible only for the gauge invariance of the theory and
interact only with gauge bosons.
\end{abstract}
\section{Introduction}

Historically, the first equation to describe the interaction of
the 1/2-spin particle with external electromagnetic field has been
nonrelativistic Pauli equation
\begin{equation}\label{eq0}
p_0\varphi (x)=\left[ \dfrac{\left( \vec p-e\vec A(x)\right) ^2}{2m}-\dfrac{%
e\vec \sigma \vec B}{2m}+eA_0(x)\right] \varphi (x).
\end{equation}
In relation (1) and below the system of units with $\hbar = c = 1$
is used; \textit{x, p, A} are 4-vectors; as usually, $xy = x^{\mu}
y_{\mu}  = x^{0}y^{0} - x^{k}y^{k}$, $\mu =0,1,2,3$, $k=1,2,3$;
$p^{\mu}  = i\dfrac{{\partial} }{{\partial x_{\mu} } }$; $\vec {B}
= rot\vec {A}$;\lb $\sigma ^{k}$ are Pauli matrices; $\varphi
\left( {x} \right)$ is a two-component wave function.

Afterwards Dirac derived his famous relativistic equation
describing the 1/2-spin particle motion. In the case of the
interaction with electromagnetic field the Dirac equation is
\begin{equation}
\label{eq1} p_{0} \psi _{D} \left( {x} \right) = \left[ {\vec
{\alpha} \left( {\vec {p} - e\vec {A}\left( {x} \right)} \right) +
\beta m + eA_{0} \left( {x} \right)} \right]\psi _{D} \left( {x}
\right).
\end{equation}
Here $\psi _{D} (x$) is a four-component wave function; $\alpha
^{i} = \left( {\begin{array}{cc}
 0 &\sigma ^{i} \\
 \sigma ^{i}&0
 \end{array}}\right)$, $\beta=\left( {\begin{array}{cc}
 I &0 \\
 0 &-I
 \end{array}}\right)$ are Dirac matrices. Unlike
equation (\ref{eq0}), the Dirac equation is linear in impulse
components $p^{\mu} $ and can be straightforwardly written in its
covariant form. The Pauli equation is a nonrelativistic limit of
the Dirac equation for the upper components of wave function $\psi
_{D} (x)$.

In principle, the relativistic generalization of the Pauli equation could be
performed in another way, with using the relativistic particle
energy-to-momentum ratio as a basis in the absence of any external fields.
This was actually done by Foldy and Wouthuysen in their classical paper [1].
The Foldy-Wouthuysen equation for free motion is
\begin{equation}
\label{eq2} p_{0} \psi _{FW} \left( {x} \right) = \left( {H_{0}}
\right)_{FW} \psi _{FW} \left( {x} \right) = \beta E\psi _{FW}
\left( {x} \right).
\end{equation}
In expression (\ref{eq2}) $E = \sqrt {\vec {p}^{\,2} + m^{2}} $.

Solutions to equation (\ref{eq2}) are plane waves of positive and
negative energies,
\begin{equation}
\label{eq3}
 \psi _{FW}^{\left( { +}  \right)} \left( {x,s} \right) = \dfrac{{1}}{{\left(
{2\pi}  \right)^{3/2}}}U_{s} e^{ - ipx};\q
 \psi _{FW}^{\left( { -}  \right)} \left( {x,s} \right) = \dfrac{{1}}{{\left(
{2\pi}  \right)^{3/2}}}V_{s} e^{ - ipx};\q p_{0} = \left( {\vec
{p}^{\,2} + m^{2}} \right)^{1/2},
\end{equation}
\noindent where $U_{s} = \left( {\begin{array}{l}
 {\chi _{s}}  \\
 {0}
 \end{array}}\right)$,  $V_{s} = \left( {\begin{array}{l}
 0\\
 \chi _{s}
  \end{array}}\right)$ are two-component normalized Pauli
functions. For $U_{s} $ and $V_{s} $ the relevant orthonormality
and completeness relations are valid.

Hamiltonian $\left( {H_{0}}  \right)_{FW} $ is related with free Dirac
Hamiltonian $\left( {H_{0}}  \right)_{D} $ by the unitary transformation. In
equation (\ref{eq2}), obvious asymmetry of space coordinates and time is seen,
although it is relativistically invariant by itself.

In the general case of the interaction with electromagnetic field $A^{\mu
}\left( {x} \right)$ there is no exact unitary transformation transforming
Dirac equation (\ref{eq1}) to Foldy-Wouthuysen $\left( {FW} \right)$ representation.
In that case Foldy and Wouthuysen found Hamiltonian $H_{FW} $ in the form of
a series in terms of powers of $1/m$ [1].

Blount [2] found the Hamiltonian $H_{FW} $ in the form of a series in terms
of powers of smallness of fields and their time and space derivatives. Case
[3] obtained an exact transformation in the presence of time-independent
external magnetic field $\vec {B} = rot\vec {A}$. In this case the Dirac
equation is transformed to equation
\begin{equation}
\label{eq4} p_{0} \psi _{FW} \left( {x} \right) = H_{FW} \psi
_{FW} \left( {x} \right) = \beta \sqrt {\left( {\vec {p} - e\vec
{A}} \right)^{2} - e\vec {\sigma} \vec {B} + m^{2}} \,\psi _{FW}
\left( {x} \right).
\end{equation}

In ref. [4] the author finds the relativistic Hamiltonian $H_{FW} $ in the
form of series in terms of powers of charge $e$ in the general case of the
interaction with external field $A^{\mu} \left( {x} \right)$.

In all the cases considered, the wave function equations in the
Foldy-Wouthuysen representation are of noncovariant form and their
Hamiltonians are nonlocal. In the $FW$ representation there is no relation
between upper and lower wave function components, i.e. $H_{FW} $ is in
essence a two-component Hamiltonian.

As the relativistic Hamiltonian $H_{FW} $ obtained in ref. [4] has
become available the possibility appeared to consider
quantum-field processes in the \textit{FW} representation within
the perturbation theory.

Section 2 of this paper briefly discusses quantum electrodynamics
(QED) in the Foldy-Wouthuysen representation [5], [6], [7]. This
Section introduces the conceptual apparatus to be used in Section
3, which addresses the electroweak theory and quantum
chromodynamics in the Foldy-Wouthuysen representation.

Section 3 also discusses the role of Higgs bosons as applied to the
formulation of the theory in the new representation.

The formulation of the standard model in the Foldy-Wouthuysen
representation does not require for the purposes of $SU(2)$
invariance of the theory the obligatory interaction of Higgs
bosons with fermions; in this case the Higgs bosons are
responsible only for the gauge invariance of the theory and
interact only with gauge bosons \textit{W}$_{\mu} ^{ \pm}  $,
\textit{Z}$_{\mu}  $.

\section{Quantum electrodynamics in the Foldy-Wouthuysen
representation}

In notation of ref. [4] the Dirac equation for quantized
electron-positron field in the \textit{FW} representation is
written as
\begin{eqnarray}
&&p_{0}\psi_{FW} (x) = H_{FW}\psi_{ FW }(x) = (\beta E + K_{1} +
K_{2} + K_{3} + \ldots)\psi_{ FW}(x);\nn\\ &&K_{1}\sim  e,\q
K_{2}\sim  e^{2},\q K_{3}\sim  e^{3}.
\end{eqnarray}
Feynman propagator of the Dirac equation in the Foldy-Wouthuysen
representation is
\begin{eqnarray}
\label{eq5} && S_{FW} \left( {x - y} \right) = \dfrac{{1}}{{\left(
{2\pi} \right)^{4}}}\,\int {d^{4}p\dfrac{{e^{ - ip\left( {x - y}
\right)}}}{{p_{0} - \beta E}}\,} = \nn\\ && = \dfrac{{1}}{{\left(
{2\pi}  \right)^{4}}}\,\int {d^{4}pe^{ - ip\left( {x - y}
\right)}\,\dfrac{{p_{0} + \beta E}}{{p^{2} - m^{2} + i\varepsilon}
} = } \nn\\ && = - i\theta \,\left( {x_{0} - y_{0}} \right)\;\int
{d\vec {p}\,\sum\limits_{s} {\psi _{FW}^{\left( { +}  \right)}} }
\,\left( {x,s} \right)\;\left( {\psi _{FW}^{\left( { +}  \right)}
\,\left( {y,s} \right)} \right)^{ +}  + \nn\\ && + i\,\theta
\,\left( {y_{0} - x_{0}}  \right)\;\int {d\vec
{p}\,\sum\limits_{s} {\psi _{FW}^{\left( { -}  \right)}} }
\,\left( {x,s} \right)\;\left( {\psi _{FW}^{\left( { -}  \right)}
\,\left( {y,s} \right)} \right)^{ +}
 \end{eqnarray}
Relation (\ref{eq5}) implies the Feynman rule of pole bypass;
$\theta(x_0)=\left\{
\begin{array}{ll}
1, &x_0>0;\\ 0, &x_0<0.
\end{array}\right.
$

 The integral equation for $\psi_{FW} (x)$ is
\begin{equation}
\psi_{FW}(x) = \psi_{0} (x)+\int d^{4}y S_{FW}(x-y) (K_{1} + K_{2}
+\ldots) \psi_{FW} (y),
\end{equation} \noindent where
$\psi_{0}(x)$  is a solution to the Dirac equation in the
\textit{FW} representation in the absence of electromagnetic field
$\left( {A^{\mu}  = 0} \right)$.

Expressions (\ref{eq5}), (8) can be used to formulate the Feynman
rules for writing of scattering matrix elements
\textit{S}$_{fi}$\textit{} and calculation of QED processes. In
contrast to the Dirac representation, in the \textit{FW}
representation there are infinitely many types of vertices of
interaction with photons depending on the perturbation theory
order: a vertex of interaction with one photon is correspondent
with factor $ -iK_{1\mu}$, a vertex of interaction with two
photons with factor $ -iK_{2\mu \nu} $, and so on. For convenience
the parts of interaction Hamiltonian terms $K_{1},K_{2},\ldots$
without electromagnetic potentials $A^{\mu}$, $A^{\mu}
A^{\nu},\ldots$ are denoted by $K_{1\mu},K_{2\mu},\ldots$,
respectively.

Each external fermion line is correspondent with one of functions (\ref{eq3}). As
usual, the positive-energy solutions correspond to particles, the
negative-energy ones to antiparticles. The other Feynman rules remain the
same as in the spinor electrodynamics in the Dirac representation.

When the external fermion line impulses lie on mass surface
\textit{(p}$^{2}$\textit{ = m}$^{2}$\textit{)} , a feature of the theory
under discussion is compensation of the contribution of fermion-propagator
diagrams with that of the relevant terms in the diagrams with the
higher-order vertices of the expansion in terms of powers of charge
\textit{e} [7]. Vertex operators \textit{K}$_{n}$\textit{}  are therewith
simplified significantly due to the law of conservation of energy-momentum.
In view of the aforesaid, the scattering matrix expansion in terms of powers
of \textit{e} can be performed, in which matrix elements \textit{S}$_{fi}$
will contain no terms including electron-positron propagators. In this case
the resultant relations are close in their structure to the relations of the
``old'', noncovariant perturbation theory developed by Heitler in the Dirac
representation [8]. A significant distinction of the relations in the
\textit{FW} representation from relations [8] is the absence of any
interaction between real electrons and positrons because of the matrix
structure of Hamiltonian \textit{H}$_{FW}$. In this representation the
electron-positron interaction can proceed only between real and intermediate
virtual states.

The interaction Hamiltonian terms \textit{K}$_{n}$ can only include an even
number of operators relating the initial and final positive-energy states to
the intermediate negative-energy states and vice versa.

To construct interactions of real particle-antiparticle in the
theory, additional terms should be introduced to the Hamiltonian
$H_{FW} $. A method to include the processes involving real
electron-positron pairs ensuring proper results in the calculation
of QED effects (for example, electron-positron pair annihilation
cross section) is to introduce the interaction between
positive-energy (negative-energy) states of equation (6) and
negative-energy (positive-energy) states of equation (\ref{eq6})
derived by the \textit{FW} transformation of Dirac equation
(\ref{eq1}) with negative particle mass
\begin{equation}
\label{eq6} p_{0} \psi _{1FW} \left( {x} \right) = \left[ {\beta E
+ K_{1} \left( { - m} \right) + K_{2} \left( { - m} \right) +
\ldots }\right] \psi _{1FW} \left( {x} \right).
\end{equation}

Equation (6) with the additional interaction can be written as
\begin{eqnarray}
\label{eq7} && p_{0} \psi _{FW} \left( {x} \right) = \beta  E \psi
_{FW} \left( {x} \right) + \left[ {K_{1} \left( {m, I,m} \right) +
\dfrac{{1}}{{2}}K_{2}} \right.\left( {m, I,m;\;m,I,\;m} \right) +
\nn\\ &&\left. +\dfrac{{1}}{{2}}K_{2}
 \left( {m,\gamma _{5} , - m;\; - m,\gamma _{5} ,\;m}
\right) + \ldots  \right] \psi_{FW}
 \left( x \right) + \nn\\
&&  + \left[ K_{1} \left( {m,\gamma _{5} - m} \right) +
\dfrac{{1}}{{2}}K_{2} \left( {m,\gamma _{5} , - m;\; - m, I,\; -m}
\right) + \right. \nn\\
 && + \left.\dfrac{1}{2}K_2 \left( m, I,m;\;m,\gamma _{5} ,\; - m \right) +
\ldots \right]\psi_{1FW}
 \left( x \right).
 \end{eqnarray}

In equation (\ref{eq7}) the notation of terms $K_{1},
K_{2},\ldots$ indicates the presence or absence of matrix $\gamma
_{5} $ near potentials $A^{\mu} \left( {x} \right)$ and the mass
sign on the left and on the right of fields $A^{\mu} \left( {x}
\right)$. Factor 1/2 of terms \textit{K}$_{2 }$ is introduced
because of two possible methods of the transition to the final
state of mass $ + m$. Similar to (\ref{eq7}), additional terms of
interaction with field $\psi _{FW} \left( {x} \right)$ can be
introduced to negative-mass equation (\ref{eq6}).

It should be particularly emphasized that in the context under discussion
the particle mass sign is some internal quantum number, which is not
associated by the author in this paper with a gravitational interaction
sign. Of course, the energy of negative-mass-sign particle is positive.

The considered interaction between equations (6), (\ref{eq6}) can
be formalized as follows. Introduce an eight-component field
$\Phi_{FW} \left( {x} \right)$, in which the four upper components
are solutions to equation (6) with positive mass \textit{(+m)},
while the lower components are solutions to equation (\ref{eq6})
with negative mass \textit{(-m).} In the case under discussion
solutions (\ref{eq3}) for free field are written as $$
\Phi_{FW}^{\left( { +}  \right)} \left( {x,s, + m} \right) =
\dfrac{{1}}{{\left( {2\pi}  \right)^{3/2}}}\left(
{\begin{array}{l}
 {U_{s}}  \\
 {0}
 \end{array}}\right)e^{-ip x};\eqno{(11a)}$$
$$  \Phi_{FW}^{\left( { +}  \right)} \left( {x,s, - m} \right) =
\dfrac{{1}}{{\left( {2\pi}  \right)^{3/2}}}\left(
{\begin{array}{l}
 {0} \\
 {U_{s}}
 \end{array}} \right)e^{-ip x};\eqno{(11b)}$$
$$  \Phi_{FW}^{\left( { -}  \right)} \left( {x,s, + m} \right) =
\dfrac{{1}}{{\left( {2\pi}  \right)^{3/2}}}\left(
{\begin{array}{l}
 {V_{s}}  \\
 {0}
 \end{array}}\right)e^{ip x};\eqno{(11c)}$$
$$  \Phi_{FW}^{\left( { -}  \right)} \left( {x,s, - m} \right) =
\dfrac{{1}}{{\left( {2\pi}  \right)^{3/2}}}\left(
{\begin{array}{l}
 {0} \\
 {V_{s}}
 \end{array}}\right)e^{ip x};\eqno{(11d)}$$

The extension of the orthonormality and completeness relations to eight
dimensions is quite evident.

Next, introduce matrices $8 \times 8$:
\begin{eqnarray*}
&& \beta _{1} = \left( \begin{array}{cc}
 I  & 0 \\
 0 &- I
 \end{array} \right),
\q\rho =\left(
\begin{array}{cc}
0 & \gamma _5 \\ \gamma _5 & 0
\end{array}
\right) , \\ &&\alpha ^k=\left(
\begin{array}{cc}
\alpha ^k & 0 \\ 0 & \alpha ^k
\end{array}
\right) ,\q \sigma ^k=\left(
\begin{array}{cc}
\sigma ^k & 0 \\ 0 & \sigma ^k
\end{array}
\right) ,\q \beta =\left(
\begin{array}{cc}
\beta & 0 \\ 0 & \beta
\end{array}
\right).\\
&&\beta_1^2=\rho^2=I;\q[\beta_1,\rho]_+=[\beta,\rho]_+=0;\\
&&[\beta_1,\alpha^k]_-=[\beta_1,\sigma^k]_-=[\beta_1,\beta]_-=0;\\
&&[\rho,\alpha^k]_-=[\rho,\sigma^k]_-=0.
\end{eqnarray*}
\setcounter{equation}{11}
 Having substituted $m \to \beta_1 m$, equations (6), (\ref{eq6}) with using matrices $8\times8$ can be
combined as
\begin{eqnarray}
\label{eq12} && p_{0} \Phi _{FW} \left( {x} \right) = \left[ \beta
E + K_{1} \left( {\beta _{1} m,\left( {I + \rho} \right),\beta
_{1} m} \right) + \dfrac{{1}}{{2}}K_{2} \left( {\beta _{1}
m,\left( {I + \rho}\right),\beta _{1} m;} \right.\right.\nn \\
&&\left. \beta _{1} m \left( {I + \rho}  \right),\beta _{1} m
\right) + \dfrac{{1}}{{4}}K_{3} \left( {\beta _{1} m,\left( {I +
\rho} \right),\beta _{1} m;\beta _{1} m,\left( {I + \rho}
\right),\beta _{1}} m;\right.\nn\\
 &&\left.\beta _{1} m,\left( {I + \rho}  \right),\beta _{1} m
  \right)
  + \ldots
\Bigg]\Phi_{FW}(x)
\end{eqnarray}
Equation (\ref{eq12}) contains equation (\ref{eq7}) for field
$\psi _{FW} \left( {x} \right)$ and the relevant equation for
field $\psi _{1FW} \left( {x} \right)$. Interaction $\left( {I +
\rho}  \right)A^{\mu} \left( {x} \right)$ enables coupling
solutions (11$a$) and (11$d$), (11$b$) and (11$c$), whereas there
is no coupling as previously between the other pairs of solutions
(11a) and (11$c$), (11$b$) and (11$d$) as well as (11$a$) and
(11$b$), (11$c$) and (11$d$).

The analysis suggests that coupling $\left( {I + \rho}  \right)A^{\mu} $
alongside the inclusion of the real electron-positron pair interaction
processes does not modify the physical results of QED processes [7].
Processes involving real negative-mass fermions appear in the theory.

Thus, with coupling $(I+\rho)A^{\mu} $ the theory in the $FW$
representation is symmetric about particle (antiparticle) mass
sign, however the signs of masses in the particle and antiparticle
interacting with each other must be opposite. Note that previously
the conclusion of opposite signs of particle and antiparticle was
made by Recami and Ziino [9] when analyzing a special relativity
theory and conditions of ``particle $ \leftrightarrow $
antiparticle'' reversibility.

Another possible method to include processes with real particle-antiparticle
pairs in the theory is coupling of the equations of motion for electron and
positron in field $A^{\mu} \left( {x} \right)$. That coupling is not
analyzed in this paper.
\section{
 Electroweak theory and quantum chromodynamics
in the Foldy-Wouthuysen representation.\\ Role of Higgs bosons in
the Foldy-Wouthuysen representation}

First write out the fermion and fermion-boson Hamiltonian of the
standard model in the Dirac representation, which is responsible
for free motion of the fermions and for the interaction of quarks
and leptons with photons, $W^{ \pm} $ and
\textit{Z}$^{0}$\textit{} particles, gluons, and Higgs bosons.
\begin{eqnarray}\label{eq13}
&&\!\!\!\!\!\!\!H_D=\sum_{f=\nu _e,e,u,d}\left\{ \left( \psi
_D\right) _f^{+}\left( \vec \alpha \vec \rho +\beta m_f\right)
\left( \psi _D\right) _f+eQ_f\left( \psi _D\right) _f^{+}\alpha
^\mu \left( \psi _D\right) _fA_\mu  \right. +
\nn\\&&\!\!\!\!\!\!\! + \frac{g_2}{\cos \theta _W}\left[ \left(
\psi _D\right) _f^{+}\alpha ^\mu \left( \frac{I-\gamma _5}2\right)
\left( \psi _D\right) _f\left( T_f^3-Q_f\sin ^2\theta _W\right)
\right. +\nn\\&&\!\!\!\!\!\!\!+\left. \left. \left( \psi _D\right)
_f^{+}\alpha ^\mu \left( \frac{I+\gamma _5}2\right) \left( \psi
_D\right) _f\left( -Q_f\sin ^2\theta _W\right) \right] Z_\mu
\right\} + \nn\\&&\!\!\!\!\!\!\! + \frac{g_2}{\sqrt{2}}\left\{
\left[ \left( \psi _D\right)^{+}_u\alpha ^\mu \left(
\frac{I-\gamma _5}2\right) \left( \psi _D\right) _d+\left( \psi
_D\right)^{+}_{\nu _e}\alpha ^\mu \left( \frac{I-\gamma
_5}2\right) \left( \psi _D\right) _e\right]  W_\mu ^{+}+
\mathrm{Hermit.conj.}\right\}+ \nn\\
&&\!\!\!\!\!\!\!+\frac{g_3}2\sum_{f=u,d}\left( \psi _{Df}\right)
_\alpha ^{+}\alpha ^\mu \lambda _{\alpha \beta }^a\left( \psi
_{Df}\right) _\beta G_\mu ^{a}-\sum_{f=e,u,d}\frac{m_f}\upsilon
\left( \psi _D\right)^{+} _f\beta \left( \psi _D\right) _fh.
\end{eqnarray}
 In (\ref{eq13}), $\left( {\psi _{D}}  \right)_{f} $ is the
Dirac fermion field; $A_{\mu}  $ is the electromagnetic field;
$Z_{\mu}  ,\;W_{\mu} ^{ \pm}  $ are the gauge boson fields;
$G_{\mu} ^{a}  $ are the gluon fields; \textit{h} is the neutral
Higgs boson field.

Besides, in (\ref{eq13}) $\alpha ^{\mu}  = \left\{
{\begin{array}{ll}
 1, &\mu = 0; \\
 \alpha ^{k}, &\mu = k = 1,2,3;
 \end{array}}  \right.$
 $Q_{f} $ is the fermion electric charge in the units of \textit{e};
$
T_{f}^{3} = 1/2$ for $f = \nu _{e},\;u;$ $T_{f}^{3} = - 1/2$ for
$f = e,\;d;
$
 $\theta _{W} $ is the electroweak mixing angle; $g_{2} = \dfrac{{e}}{{\sin\theta
_{W}} };$
 $g_{3} $ is the quantum chromodynamics coupling constant;
 $\lambda ^{a} $ are generators of group $SU(3)$; $m_{f} $ is mass of fermion
\textit{f} (in (\ref{eq13}) $m_{\nu _{e}}  = 0$ is assumed);
$\upsilon $ is the Higgs vacuum mean.

The Hamiltonian in (\ref{eq13}) is written out only for the first
lepton and quark family. For the second and third families it is
necessary to make appropriate substitutions $\left( {\nu _{e}
,\;e,\;u,\;d} \right) \to \left( {\nu _{\mu}  ,\;\mu ,\;c,\;s}
\right)\;$and$\;\left( {\nu _{\tau}  ,\;\tau ,\;t,\;b} \right)$ in
(\ref{eq13}) and introduce the quark mixing.

Previously, in ref. [7], the transition to the \textit{FW}
representation was discussed for the fifth term in Hamiltonian
(\ref{eq13}) responsible for charged current \textit{V-A}
interaction. It was shown that the transition to the
Foldy-Wouthuysen representation can be performed by similar
methods, like in quantum electrodynamics with the extension of the
Dirac matrices to eight dimensions and replacement of even
operator $C^{\mu}$ by operator $C^{\underline {\mu} } \left( {N}'
\right)^{\mu} $ and odd operator $N^{\mu} $ by operator
$N^{\underline {\mu} } \left( {C}' \right)^{\mu} $ in the relevant
expressions.\footnote{Hereinafter the even (old) operators are the
ones that do not couple (couple) the upper and lower components of
fermion fields $\psi_{FW}$.}

  The operators $\left( {{C}'} \right)^{\mu} ,\left( {{N}'} \right)^{\mu} $
differ from the previously introduced operators $C^{\mu} ,N^{\mu}
$ in  $8\times 8$ matrix $\gamma _{5} = \left( {\begin{array}{cc}
 \gamma _{5}  &0 \\
 0 &\gamma _{5}
 \end{array}} \right)$ located near matrix
  $\left( {I +
\rho}  \right)$,
\begin{equation}
\label{eq14}
\left( {{C}'} \right)^{\mu}  = \left\{ {\begin{array}{l}
 {\left( {{C}'} \right)^{0} = R\left( {\left( {I + \rho}  \right)\gamma
_{5} - L\left( {I + \rho}  \right)\gamma _{5} L} \right)R}; \\
 {\left( {{C}'} \right)^{k} = - R\left( {L\alpha ^{k}\left( {I + \rho}
\right)\gamma _{5} - \alpha ^{k}\left( {I + \rho} \right)\gamma
_{5} L} \right)R};
 \end{array}} \right.
\end{equation}
\begin{equation}
\label{eq15}
\left( {{N}'} \right)^{\mu}  = \left\{ {\begin{array}{l}
 {\left( {{N}'} \right)^{0} = R\left( {L\left( {I + \rho}  \right)\gamma
_{5} - \left( {I + \rho}  \right)\gamma _{5} L} \right)R}; \\
 {\left( {{N}'} \right)^{k} = - R\left( {\alpha ^{k}\left( {I + \rho}
\right)\gamma _{5} - L\alpha ^{k}\left( {I + \rho} \right)\gamma
_{5} L} \right)R}.
 \end{array}}  \right.
\end{equation}
In (\ref{eq14}), (\ref{eq15}), like previously, $L = \dfrac{{\beta
\vec {\alpha} \vec {p}}}{{E + m}};\;R = \sqrt {\dfrac{{E +
m}}{{2E}}} $.

Of course, the final results of calculations of specific processes
with the charged weak\lb \textit{V-A} interaction in the
\textit{FW} representation are the same as those in the Dirac
representation [7].

In a more general case of Hamiltonian (\ref{eq13}) the transition
to the Foldy-Wouthuysen representation can be also performed using
methods developed for quantum electrodynamics with the Dirac
matrix extension to eight dimensions. As a result, in the
Foldy-Wouthuysen representation Hamiltonian (\ref{eq13}) can be
written as follows:
\begin{equation}
\label{eq16} H_{FW} = \sum\limits_{f = v_{e} ,e,u,d} {\left( {\Phi
_{FW}}  \right)_{f}^{ +}  \left( {\beta E_{f} + {K}'_{1} +
{K}'_{2} + \ldots} \right)\left( {\Phi _{FW}}  \right)_{f}}
\end{equation}
In (\ref{eq16}), $E_{f} = \left( {\vec {p}_{f}^{\,2} + m_{f}^{2}}
\right)^{1/2}$ is the fermion \textit{f} free motion energy
operator. Expansion (\ref{eq16}) is in terms of powers of coupling
constants $e,g_{2}$, $g_{3}$, $\dfrac{{m_{f} }}{{\upsilon} }$ and
their reciprocal products.

The operators ${K}'_{1} ,{K}'_{2}, \ldots $ are similar in their
structure to the operators $K_{1} ,K_{2}, \ldots $ in quantum
electrodynamics with $e\alpha ^{\mu} A_{\mu}  $ replaced by
\begin{eqnarray}\label{eq17}
&& eQ_{f} \alpha ^{\mu} A_{\mu}  + \dfrac{{g_{2}} }{{\cos\theta
_{W}} }\left[ \left( {T_{f}^{3} - Q_{f}  \sin^{2}\theta _{W}}
\right)\;\alpha ^{\mu} \left(\dfrac{I-\gamma_5}{2}\right)- Q_{f}
 \sin^{2}\theta _{W} \alpha ^{\mu} \left( \dfrac{I + \gamma
_{5}} {2} \right)\right]Z_{\mu} +\nn\\ && + \dfrac{{g_{2}}
}{{\sqrt {2}} }\left\{ {\left[ {\left( {f = u} \right)\;\alpha
^{\mu} \left( {\dfrac{{I - \gamma _{5}} }{{2}}} \right)\left( {f =
d} \right) + \left( {f = \nu_{e}} \right)\;\alpha ^{\mu} \left(
{\dfrac{{I - \gamma _{5}} }{{2}}} \right)\left( {f = e} \right)}
\right]W_{\mu} ^{ +} }\right. + \nn\\ &&+\mathrm{Hermit.conj.}
\Bigg\} + \dfrac{{g_{3}} }{{2}}\left[ {\left( {f = u,d}
\right)_{\alpha} \alpha ^{\mu}  \lambda _{\alpha \beta} ^{a}
\left( {f = u,d} \right)_{\beta}  G_{\mu}^{a}} \right] -
\dfrac{{m_{f}} }{{\upsilon} }\beta h.
 \end{eqnarray}

Besides, given expressions with operators $\left({\dfrac{{I \pm
\gamma _{5} }}{{2}}}\right) $ in the Dirac Hamiltonian, even
operators $C^{\mu} $ should be replaced by operators $C^{\mu}  \pm
\left( {{N}'} \right)^{\mu} $ and odd operators $N^{\mu} $ by
operators $N^{\mu} \pm \left( {{C}'} \right)^{\mu }$ in the
relevant expressions in the\textit{ FW} representation as against
the quantum electrodynamics case.

In (17) notations $\left( {f = u} \right),\;\left( {f = d} \right)$, etc.
imply that spinor \textit{FW}-fields of associated fermions will be located
at specified places in the Hamiltonian.

By their construction Hamiltonians (\ref{eq13}) and (\ref{eq16})
are invariant under $SU(3) \times SU(2) \times\lb \times U(1)$
transformations. Note that in (\ref{eq16}) the Hamiltonian of free
motion is invariant under $SU(2)$ symmetry as opposed to the Dirac
Hamiltonian of free motion.

In their structure, in Dirac free Hamiltonian, terms $\sim\vec
{\alpha} \,\vec {p}$ and $\sim \beta m$ are transformed in
different ways in $SU(2)$ transformations, whereas in
Foldy-Wouthuysen free Hamiltonian $\sim \beta \sqrt {\vec
{p}_{f}^{\,2} + m_{f}^{2}}  $ both the subduplicates are
transformed identically in the $SU(2)$ transformations. This can
be shown straightforwardly\textit{.}

In fact, for example, the left projection operator $\left( {P_{D}}
\right)_{L} = \dfrac{{I - \gamma _{5}} }{{2}}$ in the \textit{FW}
representation is $\left( {P_{FW}}  \right)_{L} =
\dfrac{{1}}{{2}}\left( {I - \dfrac{{\beta \vec {\sigma} \;\vec
{p}}}{{E}} - \dfrac{{\beta _{1} m}}{{E}}\gamma _{5}}  \right)$;
the even part of the operator is $\left( {P_{FW}}  \right)_{L}^{e}
=\lb= \dfrac{{1}}{{2}}\left( {I - \dfrac{{\beta \vec {\sigma}
\;\vec {p}}}{{E}}} \right)$; similarly, the even part of the right
projection operator is $\left( {P_{FW}}  \right)_{R}^{e} =\lb=
\dfrac{{1}}{{2}}\left( {I + \dfrac{{\beta \vec {\sigma} \;\vec
{p}}}{{E}}} \right)$. Hamiltonian of free motion in the
Foldy-Wouthuysen representation can be represented as
\begin{eqnarray}
\label{eq18} &&\!\!\!\!\! \psi _{FW} ^{ +}  \beta E\psi _{FW} =
\psi _{FW}^{ +}  \left[ {\left( {P_{FW}}  \right)_{L}^{e} \beta
E\left( {P_{FW}}  \right)_{L}^{e} + \left( {P_{FW}}
\right)_{R}^{e} \beta E\left( {P_{FW}} \right)_{R}^{e} +
\dfrac{{1}}{{2}}\dfrac{{m^{2}}}{{E^{2}}}\beta E} \right]\psi _{FW}
= \nn\\ && = \psi _{FW}^{ +}  \left[ {\left( {P_{FW}}
\right)_{L}^{e} \beta E\left( {1 +
\dfrac{{1}}{{2}}\dfrac{{m^{2}}}{{E^{2}}} +
\dfrac{{1}}{{4}}\dfrac{{m^{4}}}{{E^{4}}} + \ldots}  \right)\left(
{P_{FW}} \right)_{L}^{e} +}  \right. \nn\\ && + \left. {\left(
{P_{FW}^{e}}  \right)_{R}  \beta E\left( {1 +
\dfrac{{1}}{{2}}\dfrac{{m^{2}}}{{E^{2}}} +
\dfrac{{1}}{{4}}\dfrac{{m^{4}}}{{E^{4}}} + \ldots}  \right)\left(
{P_{FW}} \right)_{R}^{e}}  \right]  \psi _{FW} = \nn\\ && = \left(
{\psi _{FW}^{ +} }  \right)_{L} \dfrac{{\beta E}}{{1 -
\dfrac{{1}}{{2}}\dfrac{{m^{2}}}{{E^{2}}}}}\left( {\psi _{FW}}
\right)_{L} + \left( {\psi _{FW}^{ +} }  \right)_{R} \dfrac{{\beta
E}}{{1 - \dfrac{{1}}{{2}}\dfrac{{m^{2}}}{{E^{2}}}}}\left( {\psi
_{FW}}  \right)_{R}.
\end{eqnarray}
In (\ref{eq18}) $\left( {\psi _{FW}}  \right)_{L} = \left(
{P_{FW}} \right)_{L}^{e}
 \psi _{FW} $; $\left( {\psi _{FW}}  \right)_{R} = \left( {P_{FW}}
\right)_{R}^{e}  \psi _{FW}$. From (\ref{eq18}) one can see the
desired invariance of the Hamiltonian. In view of the aforesaid an
interesting observation can be made.

The mass term in the first term of Hamiltonian (\ref{eq13}) and
the last term in (\ref{eq13}) appeared from the Higgs mechanism of
the spontaneous break of symmetry. Introduction of the mass term
to (\ref{eq13}) without the Higgs mechanism would break the
$SU(2)$ symmetry of the standard model in the Dirac
representation.

Nevertheless, regardless of the break of \textit{SU}(\ref{eq1})
symmetry, introduce the mass term to (\ref{eq13}) without the
Higgs mechanism and then transfer to the Foldy-Wouthuysen
representation. As a result we will obtain Hamiltonian
(\ref{eq16}) invariant under $SU(3) \times SU(2)\times\lb\times
U(1)$ transformations, but without the terms responsible for the
interaction of fermions with scalar Higgs bosons. Thus, the
formulation of the standard model in the Foldy-Wouthuysen
representation requires no obligatory interaction of Higgs bosons
with fermions for the purpose of the $SU(2)$ invariance of the
theory. In this case the Higgs boson sector gets narrower
significantly: the Higgs bosons are responsible only for the gauge
invariance of the theory and interact only with the gauge bosons
$W_{\mu} ^{ \pm} \;,\;Z_{\mu} .$
\section{Conclusion}

The consideration of the interacting field theory variations in
the Foldy-Wouthuysen representation allows extraction of some new
physical consequences as against the similar theory variations in
the Dirac representation. When including interactions of real
particles with antiparticles, fermions of negative mass sign (but
positive energy) appear in the theory. The theory is symmetric
about mass sign, but the particle and antiparticle masses should
be of opposite signs. In the theory there is a possibility to
relate break of $C P$ symmetry to break of symmetry in particle
(antiparticle) mass sign [7].

Finally, the standard model in the Foldy-Wouthuysen representation
with preserved $SU(3)  \times SU(2) \times  U(1)$ invariance can
be formulated without the requirement of the interaction of Higgs
bosons with fermions. Hence it appears that further theoretical
studies of the Foldy-Wouthuysen representation and comparison of
their results to available experimental data are needed.
\
\end{document}